\documentclass[12pt]{article}
\usepackage{latexsym}
\usepackage{pict2e}
\usepackage{color}
\usepackage{soul}

\begin{document}
\begin{titlepage}
\begin{center}
{\bf {Equation of State for Tachyon Neutrinos in Cosmology}}\\
\vskip 2cm
Charles Schwartz \\
Department of Physics, University of California \\
Berkeley, California 94620 \\
schwartz@physics.berkeley.edu \\
June 19, 2025 
\end{center}
\vskip 0.5cm

 Keywords: Cosmology; neutrinos; tachyons; Dark Energy; Dark Matter; Cosmic Expansion
 
 \vskip 0.5cm
 Submitted to International Journal of Modern Physics A \\
 Earlier Version Posted on arXiv:2503.19934v1
 \vskip 0.5cm

\begin{abstract}
This paper reports improvements and new results for a well developed theory of tachyons (faster-than-light particles) and follows the suggestion that neutrinos - especially the low energy neutrinos in the Cosmic Neutrino Background - may be tachyons. Starting with an application of the Thomas-Fermi method for studying a mass of particles obeying the Pauli exclusion principle, we find the following (and unusual) equations of state:
$p \sim n^{1/2} \sim \rho^{1/3}$ , where n is the particle number density, $\rho$ is the energy density, and $p$ is the pressure.

We then use these relations in the FLRW model for the evolution of the universe, focusing on the dominant role of the tachyon pressure. The result is a simple formula for the scale factor a(t); and this is found to provide an excellent fit to the famous observational data plotting "Luminosity distance" vs redshift from Type Ia supernovae. There are no free parameters in this theory; and the major uncertainty is an estimate of the tachyon mass at $0.1 eV/c^2$.

With successful quantitative explanations for both Dark Energy and Dark Matter, this tachyon-neutrino theory is presented as a strong alternative to the currently popular $\Lambda CDM$ theory. 
\end{abstract}
\end{titlepage}

\section{Introduction}
A consistent theory of tachyons (faster than light particles) \cite{CS1} developed over recent years offers remarkable success in explaining some of the great mysteries of Cosmology: Dark Matter and Dark Energy. That theoretical work starts and stays within the Special Theory of Relativity - Einstein's great theory of symmetry throughout space and time -  using both classical and quantum formulation of particle kinematics. The essential starting point for those studies was the recognition that low energy tachyons would produce large gravitational fields via the space components of the energy-momentum tensor, which is the source in Einstein's General Theory of Relativity. My most recent paper \cite{CS2} proposes two sets of experiments that may test this theory in new ways; and that paper also debunks the persistent false argument that claims tachyons are not allowed to exist.

One earlier paper \cite{CS3} looked at statistical mechanics for tachyons, but left certain issues not explored.  This paper sets out to construct the Equation of State (EOS) for the fluid of neutrinos - here assumed to be low energy tachyons - that were left over from the Big Bang and called the Cosmic Neutrino Background (CNB) in standard Cosmological theory. The results obtained in Section 2 relate particle number density $n$, energy density $\rho$, and pressure $p$ as follows.
\begin{equation}
p \sim n^{1/2} \sim \rho^{1/3}. \label{a1}
\end{equation}
These formulas are uniquely different from what we are used to in studying ordinary (slower than light) particles or radiation. That is because the basic kinematics of low energy tachyons is very different from that of the more familiar particles. Section 3 is a broad description of the new theory of tachyon-neutrinos in the cosmos.

We then present two significant applications of these EOS formulas. One is resolving a problem, noted in the appendix of reference \cite{CS2} concerning the model for Dark Energy. The other, in Section 4,  is a study of the Friedman-Lemaitre-Robertson-Walker (FLRW) theory for evolution of the universe, to see if this tachyon-neutrino theory can successfully replace the currently popular $\Lambda C D M$ model. The answer is YES. 

\subsection{Health Advisory}

It has come to the author's attention that there is a widespread mental malady among my co-professional physicists, involving a number of false and fabricated stories arguing that tachyons do not or cannot exist.
These anti-tachyon myths are simply adopted by many as being true, without the exercise of critical thinking.
Prominent among those myths are: if tachyons exist, they would lead to an intolerable causal paradox; the negative energy states of tachyons would lead to an unstable collapse of the universe; prior experiments have shown that neutrinos are not tachyons.

In my previously published papers I have examined these, and other, anti-tachyon myths; I have exposed their faults and fully debunked those claims. Reference \cite{CS1} is a good place to start if you want to read all the details.; and the most recent reference \cite{CS2} has a full disclosure on the subtle relationship between the 4-vector $k^\mu$ used to distinguish plane waves and the 4-vector $p^\mu$ used to describe relativistic classical particles.

As for experimental data: the famous OPERA experiment concluded with an upper limit for the mass of any possible tachyon neutrino; but that limit is so high that it is irrelevant for my theory. The very recent report of data from the KATRIN experiment, which examines the electron energy spectrum from the beta decay of Tritium, is rather interesting. The abstract of their latest paper \cite{KAT} gives two sets of numbers:
the upper limit on the mass of the relevant neutrino is stated as $0.45eV$ with a confidence level of $90\%$;
and a "best fit value" to all the data is stated as $m^2 = -0.14 \pm 0.14 \; eV^2$. A true negative value of $m^2$ means this particle is a tachyon. The uncertainty quoted here is such that one cannot say this proves that the neutrino is a tachyon. However it does let one draw some probabilistic inference. If one assumes a normal probability distribution (or a Poisson distribution), then one has 68\% of the total probability within the breadth of one standard deviation. The remaining 32\% of the probability is evenly split between the two wings of the distribution.  Thus, one may infer, the experiment yields an 84\% probability that this neutrino is a tachyon. 

(In their detailed analysis of the data, the latest  paper from KATRIN \cite{KAT} refers to the data points showing $m^2 < 0$ as "unphysical". [See their Section 4.4] There is nothing unphysical about tachyons. Does that language imply an improper prejudice in their analysis of the data?)

In reading reports from some reviewers selected by the editors of another journal I have found some elementary confusions about the basic math and physics of tachyon theory. I include some remarks here for the benefit of any similarly confused readers.

Some people have the notion that all you have to do is replace the mass m by an imaginary mass im, to get from theory of slower-than-light particles to a theory of tachyons.  This little trick works in some secondary or derived formulas, such as $ E = \frac{m c^2}{\sqrt{1-v^2/c^2}} \longrightarrow   \frac{i\;m c^2}{\sqrt{1-v^2/c^2}}  =  \frac{m c^2}{\sqrt{v^2/c^2-1}}.$
However this can be quite misleading as a general rule. When you write down a relativistic wave equation, Klein-Gordon or Dirac, it seems that you have replaced m by im. But the new equations imply very different boundary conditions, they lead to distinctly different mass-shells for the parameters of plane wave solutions.
In my earlier papers I found surprising new behaviors for solutions of the tachyon-Dirac equation: the helicity is Lorentz invariant; the energy-momentum tensor can have new plus-or-minus signs compared to the ordinary (slower-than-light) Dirac equation. These mathematical discoveries lead to the powerful claims that the tachyon-neutrino theory can explain both Dark Matter and Dark Energy.

One reviewer tried to find inconsistencies in the tachyon-neutrino theory by going to the "rest frame" of the particle. But tachyons do not have a rest frame: $\omega^2 -k^2 = -m^2$ does not have a solution for $k=0$.

\section{Thomas-Fermi method}
The Thomas-Fermi method is an approximation scheme that bridges classical and quantum mechanics to describe a large collection of Fermionic particles at very low energy ("zero temperature"). It was invented to describe electrons in an atom, and later applied to neutrons in a star close to collapse. To be considered a good approximation one needs many particles with very little or no interactions among them. Neutrinos in the CNB seem ideally suited to those conditions.

In any small region of space we treat the particles as a collection of free particles obeying Fermi statistics. At the lowest overall energy for this system, N particles in some volume V, we assign particles in the lowest energy states (counted in momentum space) and then sum them up. (We use units such that $\hbar = c = 1$.)
\begin{equation}
N = \frac{g V}{(2\pi)^3} \int d^3k, \;\;\;\;\; k_{min} < k  < k_{max}. \label{b1}
\end{equation}
Here g is the multiple of distinct particles: g=2 for electrons or neutrons, counting two spin states; g=12 for our neutrinos, 3 lepton types ($e, \mu, \tau$), 2 helicity states and 2 to count "particles" and "anti-particles". For ordinary particles the range of the k integral is $0 \le k \le k_F$. For tachyons it is different:
\begin{equation}
\omega^2 - k^2 = -m^2, \;\;\;\;\; \int d^3 k \rightarrow 4\pi \int_0^{\omega_F} k \omega d\omega \rightarrow 4\pi m \int_0 ^{\omega_F} \omega\; d\omega . \label{b2}
\end{equation}
That last step was the approximation for low energy tachyons $\omega << m : k \approx m$.

Here is our first result, for the particle number density n.
\begin{equation}
n = N/V = \frac{g m}{4\pi^2}\; \omega_F^2. \label{b3}
\end{equation}
To calculate the energy density we put a factor $\omega$ under the above integral.
\begin{equation}
\rho = \frac{g m}{6 \pi^2}\; \omega_F^3. \label{b4}
\end{equation}
To calculate the pressure we put a factor $v^2 =  (k/\omega)^2$ under the last integral.
\begin{equation}
p = \frac{g m^3}{2\pi^2}\; \omega_F. \label{b5}
\end{equation}

We summarize these three formulas by eliminating the quantity $\omega_F$:
\begin{equation}
p \sim n^{1/2} \sim \rho^{1/3} \label{b6}
\end{equation}
These are the equations of state for low energy tachyons obeying Fermi statistics.

We can write more precisely,
\begin{equation}
p = 3 (g/6\pi^2)^{2/3}  m^{8/3} \rho^{1/3}. \label{b7}
\end{equation}
This looks strange so lets check dimensions here. The mass m has dimensions of energy - $mc^2$ - and it also has dimensions of inverse length - $(\hbar/mc)^{-1}$. Thus both $\rho$ and $p$ scale with m as $m^4$ (energy per volume). The above equation is consistent with this.

Those relations  found above among the particle density, energy density and pressure are new and different from what physicists have seen before. That is because low energy tachyons follow kinematic rules that are unlike what we are used to with ordinary particles and radiation. Was the Thomas-Fermi method essential in achieving those relations?
I think not. If we follow conventional analysis in statistical mechanics we would write, instead of Eq.(\ref{b1}) above
\begin{equation}
n = N/V = const. \int d^3 k \frac{1}{1+e^{\omega/kT}}. \label{b8}
\end{equation}
for a system on free Fermions at temperature T. For the energy density $\rho$ and the pressure $p$ we would insert those same factors $\omega, k^2/\omega$ into this integral. In the low energy regime, we would write $k \approx m$ along with $k dk = \omega d\omega$ to get the following results.
\begin{equation}
n \sim (kT)^2,\;\;\;\;\; \rho \sim (kT)^3, \;\;\;\;\; p \sim (kT)^1. \label{b9}
\end{equation}
This leads to the exact same EOS relations we saw above, Eq.(7). In my earlier paper \cite{CS3} I studied statistical mechanics for tachyons and used the integrals shown above (\ref{b8}).  But I did not focus on the EOS relations Eq.(\ref{b6}), which will prove to be most useful in the models treated in later sections of this paper.

\section{The new model} 

In the full study of quantum theory of tachyons it emerged that there are two types, carrying opposite signs of the "gravitational charge". (That is, the energy momentum tensor has a plus-or-minus sign depending on some internal quantum number.) I assume that the wealth of neutrinos created in the Big Bang included equal numbers of all types of tachyons; thus their net contribution to gravity was zero. In that earlier epoch, nuclear forces were the dominant source of neutrino interactions; but they ceased when the cosmic temperature dropped below about 1 MeV. Eventually (and I have not seen how to estimate when this could happen), as things cooled, the gravitational forces became the dominant shaper of the tachyon neutrinos. Each type was attracted to its same members and repelled by the other type. 

To explore this idea mathematically, I constructed a Hamiltonian for a system of particles, both ordinary and tachyonic, interacting with Einstein's gravitational field (here treated in the linear approximation). The resulting formula was the starting point for the work in \cite{CS4}

The qualitative model that emerges from those previous studies is this. One type condensed, attached itself to ordinary matter collected in galaxies and the strong gravitational fields they created there produced the effects commonly ascribed to "Dark Matter". This picture was explored quantitatively and successfully in reference \cite{CS4}. The other type (carrying negative pressure) remained spread out in the rest of the universe, away from the galaxies and produced the effects commonly ascribed to "Dark Energy". This picture was presented in my earlier paper reference \cite{CS5} and we saw a quantitative fit to experimental data.

A problem was recognized in the appendix of my most recent paper \cite{CS2}: These two types of tachyon neutrinos cancelled each other out in terms of gravitational fields produced when they were all mixed up in the earlier stage of cosmic evolution; so how is that cancellation avoided by just moving some of them over here and leaving the rest over there?
The FLRW model looks at the overall summation of all types of matter in the universe.

The result found in Section 2 of this paper resolves the problem. When the one type condenses onto the galaxies (by a factor of the order of $10^3$) what happens to the pressure? Equation (7) says that it will increase at a slower rate (maybe $10^{1.5}$). Thus its contribution to the overall space-averaged pressure has decreased by this significant factor, Thus the cancellation is negligible. The original Dark Matter explanation remains intact!

That earlier paper \cite{CS3} also gave a value for the current pressure due to assumed tachyon neutrinos in the CNB that exceeds by almost one order of magnitude the value derived from conventional theory and used in my first paper \cite{CS5} explaining Dark Energy. (See Eq. (\ref{d5}) ahead.) I am unsure about the reliability of that calculation; and this indicates a modest uncertainty in the results we derive below. This uncertainty (which I hope will be reduced by further theoretical analysis) is truly small compared with the commonly recognized fact that reigning estimates of the Cosmological Constant $\Lambda$ are off by some hundred orders of magnitude.

Readers who wish to read more about the history and problems of the $\Lambda CDM$ theory are invited to  the review contained in \cite{MI}.

\section{FLRW} 
This famous model looks at the whole universe, taking all the stuff in it spread out as a homogeneous isotropic fluid described by an energy density $\rho$ and a pressure $p$. These are at first unknown functions of time. The metric has a scale factor $a(t)$ and one wants to learn how the universe expands or contracts with time. Einstein's equation leads to the following (called the Friedmann equations).
\begin{eqnarray}
\frac{\ddot{a}}{a} = \Lambda/3 - (4\pi G/3) (\rho + 3p) \label{d1} \\
(\frac{\dot{a}}{a})^2 = (8\pi G/3) \rho - k/a^2 + \Lambda/3. \label{d2}
\end{eqnarray}
Here $\Lambda$ is the "cosmological constant", G is the gravitational constant, and $k$ is the spatial curvature. The dot on a(t) means time derivative.

Authors usually start their work with these equations by making some assumption about which of the physical inputs are dominant, in some epoch. My focus is on tachyon neutrinos and they produce a large pressure. It is taken as obvious that the particle density, for any sort of particles, depends on a(t) as $ n \sim 1/a^3$. From the tachyon EOS derived above, we now write $p = p_0/a^{3/2}$, where the subscript 0 means the present time $t=0$ and a(0) = 1.

I will now put this formula for p into the first Friedmann equation, drop $\Lambda$ and $k$.  For low energy tachyons the energy density $\rho$ is negligibly small; but some other matter may contribute significant energy density. We can eliminate that variable $\rho$ by using the second Friedmann equation .This leads me to the following nonlinear differential equation:
\begin{equation}
\ddot{a}  + (\dot{a})^2/(2a) = q a^{-1/2}, \;\;\;\;\; q = -4\pi G p_0. \label{d3}
\end{equation}
Note that in my model the pressure $p_0$ is a negative number; so $q > 0$. This equation can be solved. 
\begin{equation}
a (t) = [1+\frac{3 H_0}{2}   t + \frac{3q}{4}t^2]^{2/3} \label{d4}
\end{equation}
$H_0$ (a constant of integration in my solution) is called the Hubble constant at the present time. At large times this result implies unlimited expansion of the universe.

I should be clear about the range of the time variable t in this solution. Positive values of t refer to the future, relative to today; negative values of t refer to the past, relative to today. In the complete theory of FLRW, $a \rightarrow 0$ at some large negative time: that is called the Big Bang. My theory, with tachyon neutrinos as the dominant feature of the universe's $T^{\mu \nu}$, is reasonable only for a select portion of that history. While I have not been able to define exactly when the CNB tachyons became low energy and exerted the large gravitational fields, it was certainly well after the earliest epochs when radiation, and then cold matter, were dominant.  Thus one should not be surprised that Eq. (14) does not take us to $a=0$. This model is offered as valid for the present and for recent epochs; and the data that we shall fit this to in the next section is exactly in that region.

It appears that we have a simple model here that does not require invocation of any mysterious "cosmological constant"; nor is Dark Matter a mystery; nor is Dark Energy a mystery.  The single assumption - that neutrinos, especially in the CNB, are tachyons with a mass about $0.1 eV/c^2$ - solves all those mysteries. Let's check some numbers.

In my 2017 paper \cite{CS5} I gave an estimate of the negative pressure of tachyon neutrinos in the CNB as
\begin{equation}
p_0 \approx  6,800\;eV/cm^3. \label{d5}
\end{equation}
This was based upon conventional BB theory that gives the number density of CNB  as $340/cm^3$;
I also used the conventional theory for the average kinetic energy of the CNB as $1.7 \times 10^{-4}$ eV per particle. My subsequent paper \cite{CS3} explored statistical mechanics for tachyons, as it should differ from statistical mechanics for ordinary particles; and my conclusions there were that this estimate needs a modest increase

The one important parameter in my model is q, which has the dimensions of $time^{-2}$

\begin{equation}
q = 4 \pi G p_0/c^2  \label{d6}
\end{equation}

Lets get some familiar constants; start with the critical density that is defined  in conventional cosmology:
\begin{eqnarray}
\rho_{crit} = \frac{3H_0^2}{8\pi G} = 1.878 h^2 \times 10^{-29} \; g\; cm^{-3} = 1.05 \times 10^4 (eV h^2/ c^2) cm^{-3}, \label{d7}\\
g  (1 gram )=   5.61 \times 10^{32}  eV/c^2 \label{d8} \\
q = 4 \pi G p_0/c^2 = 1.5  H_0^2 \frac{p_0/h^2}{10^4 eV/cm^3} =   (0.97/h^2) H_0^2. \label{d9}
\end{eqnarray}
This  h is an experimental factor (close to $0.7$) for the Hubble constant in astronomers' units.

This all looks very comfortable. The tachyon neutrino pressure in the CNB fits neatly with the conventional lifetime of the universe ($T = H_0 ^{-1}$).  Without the $v/c \sim 600$ estimate for the tachyons this calculation would look  quite different. 

\section{Fitting the famous data} 
The crux of the "Dark Energy" mystery is the analysis of experimental data collecting light signals from Type Ia supernovae. These are considered to be "standard candles" throughout the universe. The particular curve of interest plots the "luminosity distance" ($d_L$) of each source vs the redshift ($z$) of the light received from that source. The observable redshift (z) is an indicator of the distance of the source. The theoretical link of the Robertson-Walker metric to these variables I take from Chapter 1 of Weinberg's classic book, "Cosmology" (2008) \cite{SW}. Here are the relevant equations.
\begin{eqnarray}
1 + z(t) = 1/a(t), \label{e1}\\
d_L (z) = (1+z)\int_{t_1}^0 dt/a(t). \label{e2}
\end{eqnarray}
The time $t_1 < 0$ is the time when the signal was sent by the distance source; and the time $t=0$ is the present time, when the signal is received here. My convention is $a(0) = 1$. The conventional theory (as presented in Weinberg) uses the Friedmann equations to get implicit representation of a(t) in terms of assumed sources of matter in the universe ($\rho$ and $p$). That theory admits a cosmological constant in Einstein's equation and the magnitude of that "vacuum energy" $\Lambda$ is treated as a free parameter to be fitted to the data.

My alternative theory, presented above, assumes only that neutrinos are tachyons, with a mass estimated to be about $0.1 eV/c^2$; and I have obtained, assuming that the tachyon pressure is the dominant factor, an explicit formula for a(t). The only free parameter in my theory turns out to be the Hubble constant $H_0$.  To make the calculations most simple, I will approximate the exact formula Eq.(\ref{d4}) for a(t) due to the dominance of tachyon pressure in the universe as follows.
\begin{equation}
a(t) \approx (1+ t/s)^s. \label{e3} 
\end{equation}
Here the time t is given in units of $H_0^{-1}$ and s is an as yet undetermined number. Note that my exact formula Eq.(\ref{d4}) has a quadratic form in t to the power 2/3. That quadratic is not so far from a perfect square; and that leads me to offer the approximation Eq.(\ref{e3}).

Putting this into the equations above I get the final formula,
\begin{equation}
d_L (z) = \frac{s}{s-1} [(1+z)^{2-1/s} - (1+z)]. \label{e4}
\end{equation}
Using the value $s = 4/3$ this reads,
\begin{equation}
d_L(z) = 4[(1+z)^{5/4} - (1+z)]. \label{e5}
\end{equation}
At small z this is approximated by $d_L \sim z + (5/8) z^2 + ...$; and at large z we have $d_L \sim 4 z^{5/4}$. At z=1, this is $d_L = 1.514$.
How does this compare with the experimental data? If I look at Weinberg's Figure 1.3, this seems an excellent fit to the curve he presents as the conventional theory's best fit to the experimental data. ( Again: they have two adjustable parameters, I have none.)  I will refer this comparison to others who have access to more recent data (at higher z) and experience at fitting that data to theorists' formulas.

\section{Conclusion and future work}
The new results presented in Section 2 allow us to fortify the claim that our tachyon-neutrino theory can displace the now dominant theory known as $\Lambda CDM$.  The $\Lambda$ in that acronym refers to the Cosmological constant, an old mathematical invention that was resurrected to "explain" Dark Energy. Attempts to justify this on physical grounds ("zero point energy") are off by many many orders of magnitude. As seen in Section 5, my  model fits the data quantitatively within a small factor of uncertainty. (The value of $p_0$ given in (\ref{d5}) above comes from a formula \cite{CS4} that involves the factor $m^2$.)

Previous study \cite{CS4} showed that the condensed tachyon neutrinos can explain the observed phenomena now ascribed to a mysterious Dark Matter. The application of the Thomas-Fermi model to that topic will be the focus of future work.

I invite more physicists to join in exploring this exciting (unconventional) theory.

\vskip 0.5 cm 
{\bf Acknowledgment}
I want to thank Rodrigo Verney for helpful conversations during my work on this subject.

\vskip 0.5cm
\setcounter{equation}{0}
\def\theequation{A.\arabic{equation}}
\boldmath
\noindent{\bf Appendix A:  Care with the word "pressure"}
\unboldmath
\vskip 0.5cm

Pressure, as a word, is first introduced by us physics teachers to mean Force-per-unit-area ( for example, we put our hand in front of a water hose). When we teach thermodynamics and statistical mechanics we refer to a lively gas of molecules, rapidly bouncing off the walls of their container. What do we say about a gas of neutrinos? We commonly describe our bodies here on earth being penetrated by a fierce flow of neutrinos from the sun; and we don't notice it at all. The neutrinos in the CNB are very much lower in energy and so their crossection for any sort of interaction is many orders of magnitude less.

Pressure is the word assigned to describe the flow of a mass of stuff that is not at rest. But the word "flow" suggests a direction, while this word pressure is used without any reference to a direction. Maybe it means just the average magnitude of the momenta of the particles.

This word problem gets more complicated when we talk about the energy-momentum tensor. This quantity has two roles in theoretical physics: it describes the flow (perhaps one should use the word "swarm", which describes a bunch of fast-moving bodies that don't bump into one another) in a purely kinematic description; it is also the source of gravitation in Einstein's General Relativity. In the traditional sense, these two are the same. However, in my theory of tachyons, I find that there are some plus-or-minus factors that appear in the construction of $T^{\mu \nu}$.  Thus my theory acknowledges a possible negative value for the pressure

\end{document}